\documentclass[twocolumn,prl,superscriptaddress,showpacs]{revtex4-1}

\usepackage{graphicx}
\usepackage{amsmath,amssymb}
\usepackage{epstopdf}

\begin{document}

\title{Breaking anchored droplets in a microfluidic Hele-Shaw cell}
\date{\today}

\author{Gabriel Amselem}
\affiliation{LadHyX and department of Mechanics, Ecole Polytechnique, CNRS, 91128 Palaiseau Cedex, France}
\author{P.T. Brun}
\altaffiliation{Current address: Department of Mathematics, Massachusetts Institute of Technology, Cambridge, Massachusetts 02139, USA}
\affiliation{LFMI, EPFL, Lausanne, Switzerland}
\author{Fran\c{c}ois Gallaire}
\affiliation{LFMI, EPFL, Lausanne, Switzerland}
\author{Charles N. Baroud}
\email{baroud@ladhyx.polytechnique.fr}
\affiliation{LadHyX and department of Mechanics, Ecole Polytechnique, CNRS, 91128 Palaiseau Cedex, France}

\begin{abstract}
  We study microfluidic self digitization in Hele-Shaw cells using pancake droplets anchored to surface tension traps.
  We show that above a critical flow rate, large anchored droplets break up to form two daughter droplets, one of which
  remains in the anchor. Below the critical flow velocity for breakup the shape of the anchored drop is given by an
  elastica equation that depends on the capillary number of the outer fluid. As the velocity crosses the critical value,
  the equation stops admitting a solution that satisfies the boundary conditions; the drop breaks up in spite of the
  neck still having finite width. A similar breaking event also takes place between the holes of an array of anchors,
  which we use to produce a 2D array of stationary drops {in situ}.
\end{abstract}

%47.15.gp	Hele-Shaw flows
%47.55.df	Breakup and coalescence
%47.61.Fg	Flows in micro-electromechanical systems (MEMS) and nano-electromechanical systems (NEMS)

\maketitle

One of the underlying drivers in microfluidics is to miniaturise the standard multiwell plate and transform it into an
integrated programmable device, while replacing the hundreds of wells typical today with thousands or more
nanoliter-scale compartments. Early success was achieved by using deformable chambers that could be opened or closed
using an external pressure source~\cite{thorsen02}. In parallel droplet-based microfluidics continues to attract ever
increasing interest since it provides an elegant way to encapsulate an initial sample into a large number of independent
micro-compartments. However, the traditional droplet production and manipulation methods all rely on the drops flowing
in a row in a linear microfluidic channel. Studying the contents of these drops~\cite{song03,miller12} is therefore more
akin to flow cytometry than to multiwell plates.

Recently several groups have shown how to array droplets in micro-fabricated traps, particularly in a wide
two-dimensional (2D) region~\cite{huebner09,abbyad-dangla}. These devices typically allow a droplet density of several
hundred per cm$^2$, far beyond what is currently possible outside microfluidics. These devices must still be coupled
nevertheless to a traditional drop production device before these are brought to the observation chamber. While the
underlying physical mechanisms for these drop production devices is now well understood~\cite{garstecki05,vansteijn09},
they are poorly adapted in practice to making a limited number of stationary drops. For this reason, quasi-two
dimensional devices have been designed to break an initially large drop into stationary sub-droplets that are held in
pockets on the side of a sinuous channel~\cite{boukellal09,yamada10,cohen10}, while truly 2D devices would allow much
higher density of trapped droplets~\cite{wu2012shrunk}.

In this letter we describe the ability to break droplets \textit{in-situ} in a wide chamber, by pushing them over a
truly two dimensional array of micro-fabricated traps~\cite{dangla11}. We elucidate the physical mechanisms first on a
single droplet and show that it is well described by a set of universal curves given by the elastica equation. The drop
then breaks through a singularity in the curves beyond a critical deformation, leading to a well characterized and
robust size. In addition to its applications for droplet arrays, this new route to breaking a liquid interface provides
fundamental insight into the evolution of drops and bubbles in all confined geometries, where the traditional
Rayleigh-Plateau instability is not active.

%%Figure 1
\begin{figure}[!h]
\includegraphics[width=\columnwidth]{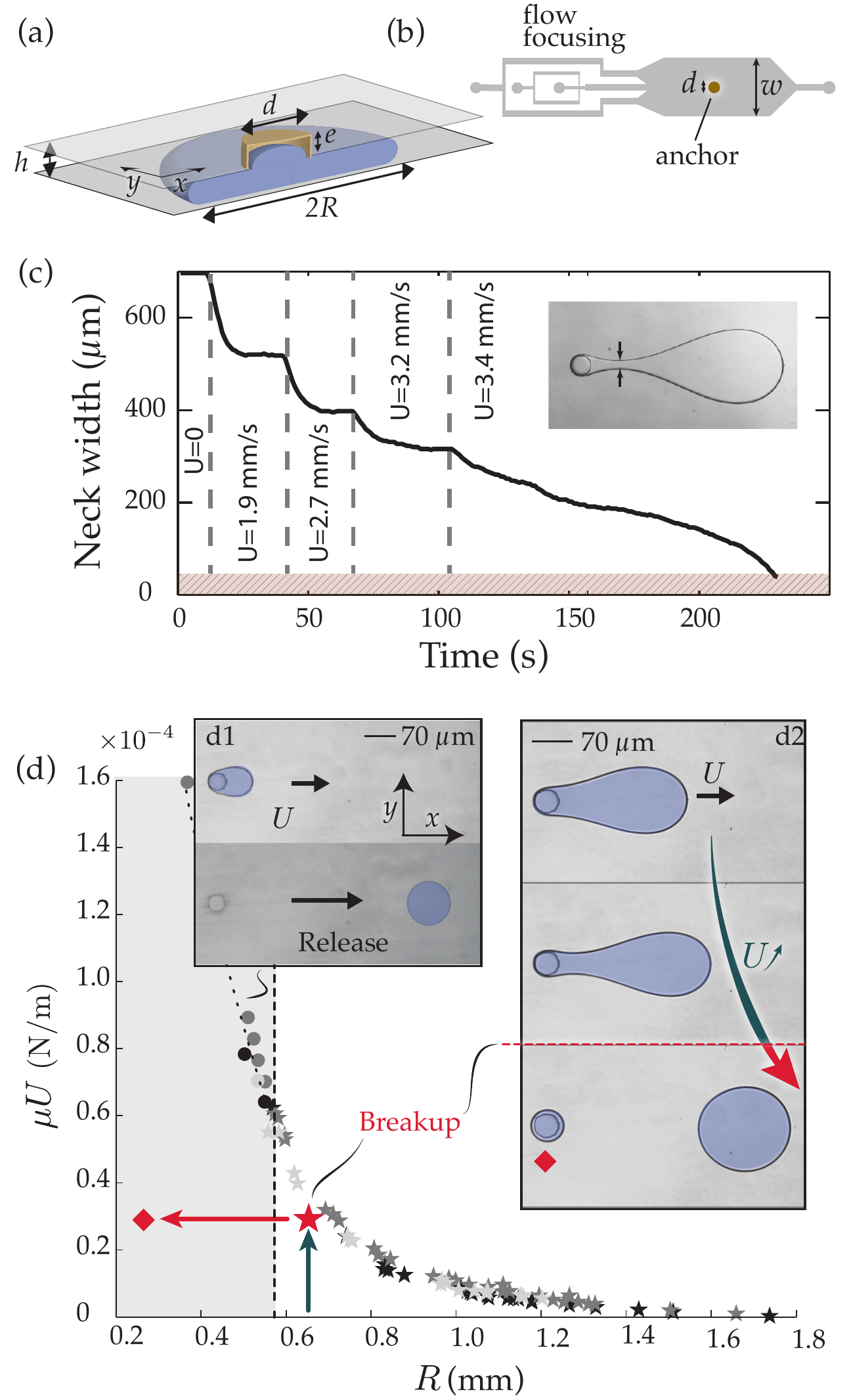}
\caption{ (a) Schematic of an anchor of diameter $d$ and height $e$, trapping a drop of radius $R$ (no external flow),
  the chamber height is $h$. (b) Sketch of the microfluidic device: droplets are produced with a flow focuser and
  trapped on an anchor centered in the observation chamber, $w=2.5$~mm. (c) Typical evolution of the droplet neck width
  (see inset) as the outer flow velocity $U$ is increased. When the critical velocity is passed the neck never reaches a
  stationary state; instead it decreases until it matches the channel height ($h=49~\rm \mu$m), where it becomes
  unstable by Rayleigh-Plateau instability. (d) For a given microchannel, a trapped droplet either escapes from its
  anchor at a critical value of the capillary number (greyed area, $R\lesssim 600~\rm \mu m$) or breaks on the anchor
  ($R\gtrsim 600~\rm \mu m$), leaving a smaller droplet behind. Experimental points for droplets of FC-40 in
  water/glycerol mixtures of different viscosities: $\mu=0.93\;\rm{cP}$ (dark grey), $\mu=1.6\;\rm{cP}$ (medium grey),
  and $\mu=3.3\;\rm{cP}$ (black). SDS was used as a surfactant at a concentration of 2\% in all cases.}
\label{fig:setup}
\end{figure}
%

%Experimental setup - geometric parameters
The experimental setup consists of wide chamber (width $w=2.5~$mm) in which the drops can be anchored in a central trap,
as shown in Fig.~\ref{fig:setup}a-b. The chamber height $h$ and the anchor diameter $d$ were in the range $35-50 \; \rm
\mu m$ and $75-200 \; \mu\rm m$, respectively. The height of the anchor $e$ is chosen to be of the order of the height
of the channel.  All devices were made out of polydimethylsiloxane (PDMS, Dow-Corning). The droplets studied here were
produced by a flow-focusing junction upstream of the chamber (Fig.~\ref{fig:setup}b), designed such that the drops were
forced to adopt a pancake shape, of radius $R\gg h$, in the chamber. Droplets of fluorinated oil were produced in a
glycerol/water mixture with 2\% SDS as a surfactant.  The oil viscosities $\mu_{\rm oil}$ ranged between 1.2 and 24~cP,
whereas the outer phase viscosity $\mu$ was varied between 0.89 and 3.3~cP by varying the water to glycerol ratio. The
interfacial tension between the drop and the outer liquid had a typical value $\gamma\sim 17\, \rm mN/m$.

%Experimental setup - what the experiment
During a typical experiment, a droplet is anchored on a surface energy trap and the outer velocity is increased
step-by-step. A typical measurement of the drop evolution is shown on Fig.~\ref{fig:setup}c, where the smallest width of the drop neck
is reported. We observe that the neck reaches a stationary value when the outer fluid is flowing below the critical
velocity. At the critical velocity, the neck begins to decrease from a finite value, slowly at first but then this
decrease is accelerated until the neck width becomes equal to the channel height. Because of this loss of
confinement locally, the Rayleigh-Plateau instability becomes active and the neck breaks, as observed previously for
different droplet production situations~\cite{garstecki05,dollet08,vansteijn09}.

For a given setup and droplet volume, we find that there exists a critical capillary number ${\rm
  Ca^\star}=\mu U^\star/\gamma$ above which the droplet either escapes from the anchor~\cite{dangla11}
(Fig.~\ref{fig:setup}-d1), or breaks on the anchor (Fig.~\ref{fig:setup}-d2). In this second regime, the drop leaves
behind a daughter droplet small enough to remain anchored at the breakup velocity $U^\star$: the daughter droplet lies
in the trapped region of the phase diagram, as indicated by the diamond in Fig.~\ref{fig:setup}d. In the rest of this
letter we focus on the breakup regime exclusively.

%%%Figure 2
\begin{figure}[!h]
\begin{center}
\includegraphics[width=\columnwidth]{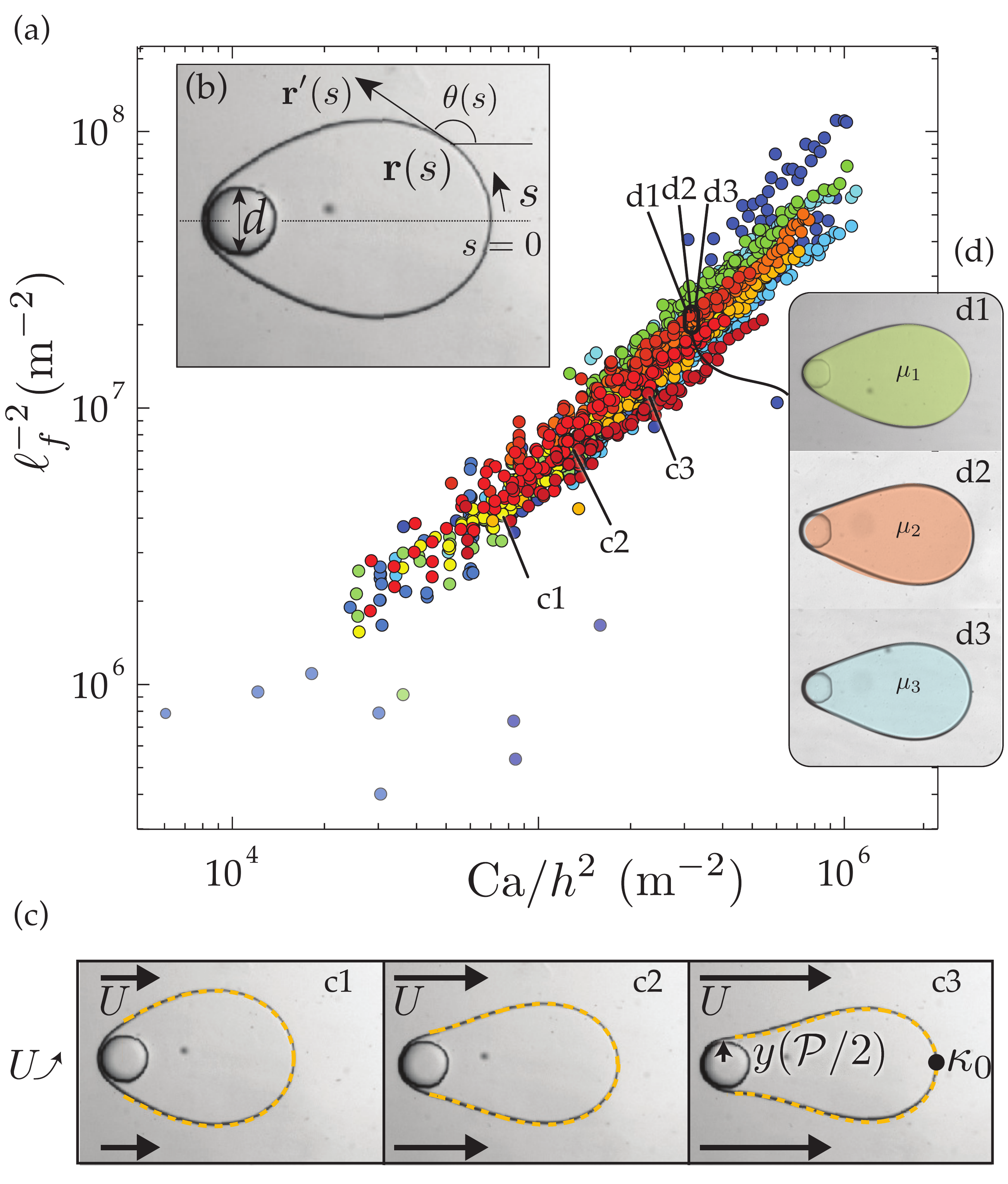}
\caption{ (a) Evolution of the fitting parameter $\ell_f$ with the value of ${\rm Ca}/h^2$, for different channel
  heights, trap diameters and droplet fluid viscosities. The cloud of data points confirms that $\ell^{-2}\sim{\rm
    Ca}/h^2$. (b) Photograph of an anchored droplet under flow and system of coordinates in which the interface $\mathbf
  r(s)$ is derived.  $s$ is the arclength along the interface and $\mathbf{r}'(s)$ is its tangent. (c)~From left to
  right : droplet shape for increasing outer fluid velocities. Pink dotted lines are the best fitting elastica
  shape. (d)~Three droplets of different viscosities ($\mu_1=1.2, \mu_2=4.1, \mu_3=24$~cP) but similar volumes, for the
  same value of $\mu U/\gamma h^2$: the droplet shape is independent of the droplet viscosity.}
\label{fig:fig2}
\end{center}
\end{figure}

%droplet shape - derivation of the equations
The droplet stationary shapes are imposed by the pressure difference between the two sides of the interface. The flat
microfluidic chamber can be modeled as a Hele-Shaw cell so we adopt a two-dimensional depth averaged formalism to
describe the system. Then, the pressure drop in the flow direction $x$ is related to the average flow velocity $U$
according to
\begin{equation}
\label{eqn:heleshaw}
p_\text{o}(x)=-24\frac{\mu U}{h^2}x + \rm{Cst},
\end{equation}
where $p_\text{o}(x)$ is the pressure in the outer, aqueous phase. The pressure $p_\text{i}$ inside the oil drop is
constant~\cite{dangla11,Nagel2014} and is related to the outside pressure $p_o(x)$ by the Laplace relation:
\begin{equation}
\label{eqn:laplace}
p_\text{i}-p_\text{o}(x)=\gamma(\kappa_{\perp} + \frac{\pi}{4} \kappa_{\parallel}),
\end{equation}
where $\kappa_{\perp}$ and $\kappa_{\parallel}$ are the curvatures in the perpendicular and the parallel planes
respectively~\cite{Park1984}. Away from the anchor, the curvature in the perpendicular plane is assumed constant
$\kappa_{\perp}=2/h$.

The droplet 2D shape $\mathbf{r}(s)$, where $s$ denotes the interface arc-length, is described in
Fig.~\ref{fig:fig2}b. The tangent to the droplet interface is obtained by differentiating the position with respect to
the arc-length $\mathbf{r'}(s)=( \cos \theta(s),\sin\theta(s))$, and the in-plane curvature is given by
$\kappa_{\parallel}=\theta'(s)$. Differentiating Eq.~(\ref{eqn:laplace}) with respect to $s$ and using
Eq.~(\ref{eqn:heleshaw}) in the particular case where $x=\mathbf{r}(s)\cdot\mathbf{e_x}$ yields an elastica equation:
\begin{subequations}
\label{eqn:system}
\begin{equation}
\label{eqn:elastica}
\theta''(\overline{s})-\frac{R^2}{\ell^2}\cos\theta(\overline{s})=0,
\end{equation}
%
% h \sqrt{\frac{\pi}{96}}\sqrt{\frac{\gamma}{\mu U}} 
where $\ell= h \sqrt{\frac{\pi}{96}}{\rm Ca}^{-1/2}$ denotes the ``visco-capillary length'' of the problem and the
radius $R$ of the undeformed droplet is used to non-dimensionalize the problem (non-dimensional variables, such as $\bar
s$, are denoted with a bar). This is a pendant drop equation~\cite{dangla11,majumdar76}.

%1st qualitative check of the equations: independence of shape and internal viscosity
Note that the droplet shape does not depend on its viscosity: the only viscosity entering the problem (through $\ell$)
is the external fluid viscosity. To qualitatively check this prediction, we show on Fig. ~\ref{fig:fig2}d the shapes of
three different droplets of different viscosities (between 0.77 to 24~cP) but similar volumes, and at the same value of
Ca. The drops indeed have similar shapes that cannot be distinguished from each other.

%Integration of the elastica
Secondary variables are now defined to ease the integration of the problem. The cumulative dimensionless area swept by
the droplet interface, $\overline{\alpha}(\overline s)$, is defined as the solution of
\begin{equation}
\label{ }
\overline{\alpha'}(\overline s)=\overline y(\overline s)\overline{ x'}(\overline s).
\end{equation}
\end{subequations} %end of the system of equation started 30 lines earlier...
%how do we obtain the droplet shapes
\noindent The drop shape is obtained by integrating Eqs. (\ref{eqn:system})a \& b with a shooting method. For given
values of $\overline d$ and $\overline \ell$, we use the curvature
$\overline{\kappa_0}=\overline{\kappa}_{\parallel}(\overline{s}=0)$ as the sole shooting parameter and search for the
values of $\overline{\kappa_0}$ that satisfy the geometric condition $\overline{y}=\overline{d}$ at
$\overline{\alpha}=\pi/2$.  The droplet shape $\mathbf{r}(s)$ is then fully defined by the triplet $\{\overline{d},
\overline{\ell}, \overline{\kappa_0} \}$.

%Method to get all the numerical shapes by varying all parameters, and comparison with experiments
The calculations of $\overline{\kappa_0}$ are first performed for different values of $\overline \ell$, while keeping
$\overline d$ constant, and the whole process is repeated for different values of $\overline d$. This leads to a catalog
of shapes that can be used to fit the experimental droplet shapes. For each experimental condition, the best fitting
shape is found and its associated triplet is called $\{\overline{d_f},\overline{\ell_{f}},\overline\kappa_{0f}\}$.
Since all numerical shapes are obtained using dimensionless variables, it is also necessary to re-dimensionalize them
and we call $R_f$ the radius of the best fitting shape for each fit. There is a very good agreement between numerical
and experimental shapes, as shown on Fig.~\ref{fig:fig2}c. We also find excellent agreement between $R_f$ and $R$, with
a difference of at most 5\% between both values, the best fits being obtained for low values of $\ell$.

%Now we get quantitative and compare $l_f^{-2}$ and Ca/h^2
A quantitative comparison between theory and experiments is obtained by comparing the evolution of the fitting parameter
$\ell^{-2}_f= R_f^2/\overline{ \ell^{2}_f}$ with the experimental value of $ {\rm Ca}/h^2$. The resulting values are
plotted in Fig.~\ref{fig:fig2}a, which displays the data for different channel geometries, droplet and outer fluid
viscosities, and trap diameters. The data verify the scaling predicted by theory $\ell^{-2}\sim {\rm Ca}/h^2$, with a
prefactor ($\sim 60$) that differs  from the prediction ($96/\pi\simeq31$), most likely because of dynamic
surfactant effects~\cite{dangla11}.  Taken together, the above results show that the droplet shape is well described
by the elastica for forcing values below the breaking threshold.

%%%Figure 3
\begin{figure}[!h]
\begin{center}
\includegraphics[width=\columnwidth]{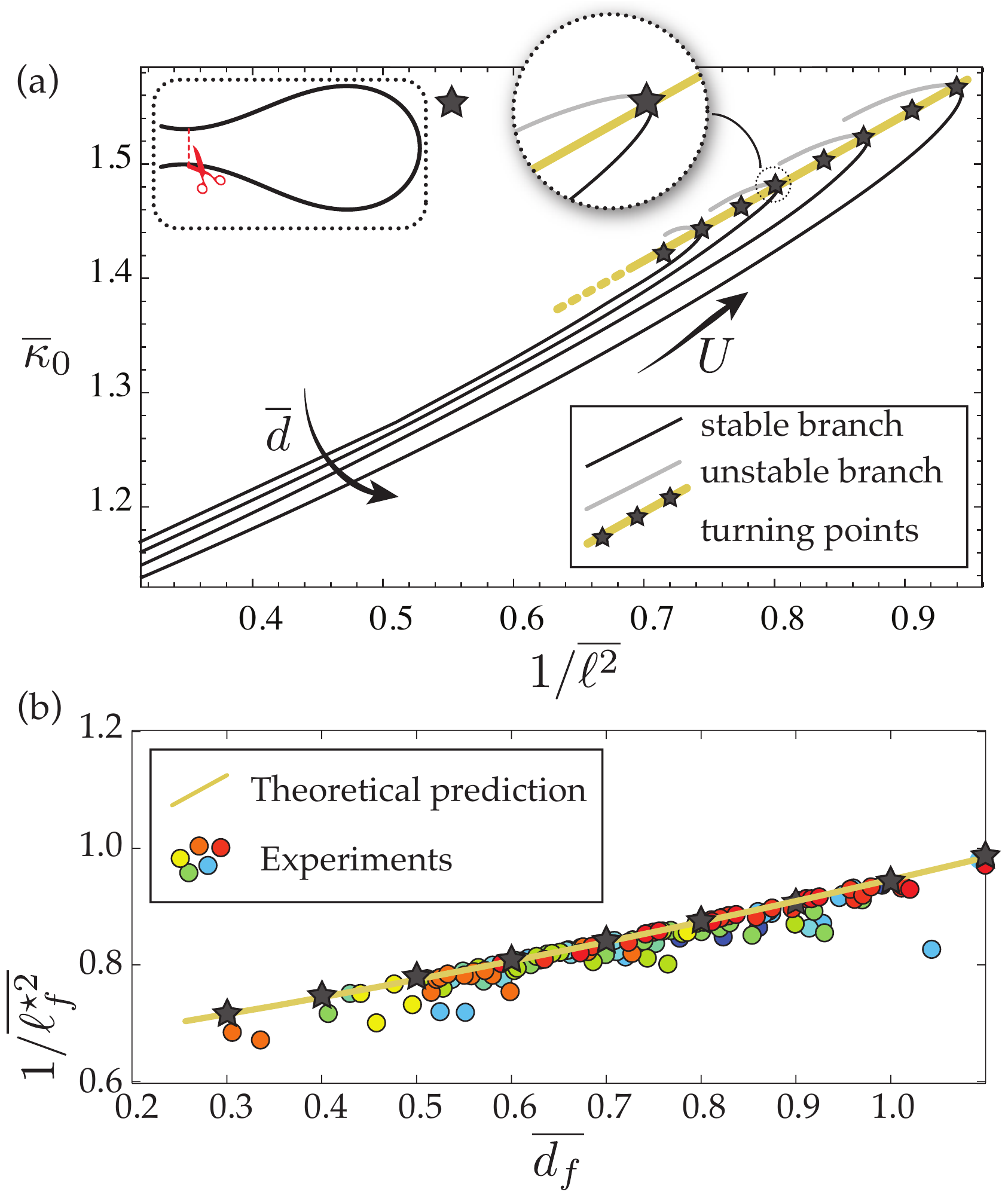}
\caption{(a) Branches of solutions of the elastica shown in the plane $\left(1/\overline
    {\ell^2},\overline{\kappa}_0\right)$, each corresponding to a given value of $\overline d$.  The stable part
  (reps. unstable) is denoted with a black line (resp. gray line). For a given $\overline d$, the critical value
  $1/\overline{\ell^{\star2}}$ at which the drop breaks is indicated by a star, corresponding to the turning point of
  the branch of solutions. Yellow line shows the interpolation of a discrete number of turning points. Inset: shape of
  the anchored drop at the critical value of $1/\overline{\ell^{\star2}}$. (b) Comparison between experimental values of
  $1/\overline {\ell^\star2}$ at which the drops break with the theoretical prediction from part (a).}
\label{fig:fig3}
\end{center}
\end{figure}

%OK the model is good for shape description, can it tell us anything about breaking?
We now turn to the breaking. The solutions to the elastica equation can be plotted as a family of curves in the
($\overline{\kappa_0}$, $1/\overline{\ell^{2}}$) plane, for different values of $\overline d$, see
Fig.~\ref{fig:fig3}a. The curves all display a monotonic increase of $\overline{\kappa_0}$ with $1/\overline{\ell^{2}}$,
until a maximum value of $1/\overline{\ell^{2}}$ where they reach a turning point, and after which no values of
$\overline{\kappa_0}$ are found. The folding of the branch of solutions is associated to an exchange of stability at the marginally stable turning point. This implies that no stationary solutions to the elastica equation can be found beyond
this point and the equilibrium can only be reached through a dynamic process which is not accounted for by the static
model.  We call $\overline{\ell^\star}$ the value of $\overline{\ell}$ at the turning point of the curve, and therefore expect  that the droplet will break  at  $\overline{\ell}=\overline{\ell^\star}$, even though the calculated 2D droplet shapes the droplet neck still has a finite width, as shown in the inset of
Fig.~\ref{fig:fig3}a.

%What happens during the experiment, how do we reach l=l^\star?
Experimentally, increasing quasi-statically the flow rate around a droplet pinned on an anchor amounts to walking along
a curve with increasing $1/\overline{\ell^2}$. This leads to an increase in $\overline{\kappa_0}$ until the value of
$\ell=\ell^\star$ is reached, corresponding to a critical velocity $U^\star$ beyond which there are no stationary
solutions. For each experiment we compare the fitted value of $\overline{\ell_f^\star}$ with the predicted value
$\overline{\ell^\star}$ in Fig.~\ref{fig:fig3}b. The agreement between the two is remarkable, confirming the
interpretation of the breaking: above the velocity $U^\star$, no equilibrium droplet shape can be found that satisfies
the elastica equation with the imposed boundary conditions. We find that the larger $\overline{d_f}$, the larger the
value of $1/\overline{\ell^\star}$, \textit{i.e.} in a given experimental setup with a fixed trap diameter $d$, the
critical velocity $U^\star$ leading to break-up is smaller for large droplets than it is for small droplets.

\begin{figure}[!h]
\begin{center}
\includegraphics[width=\columnwidth]{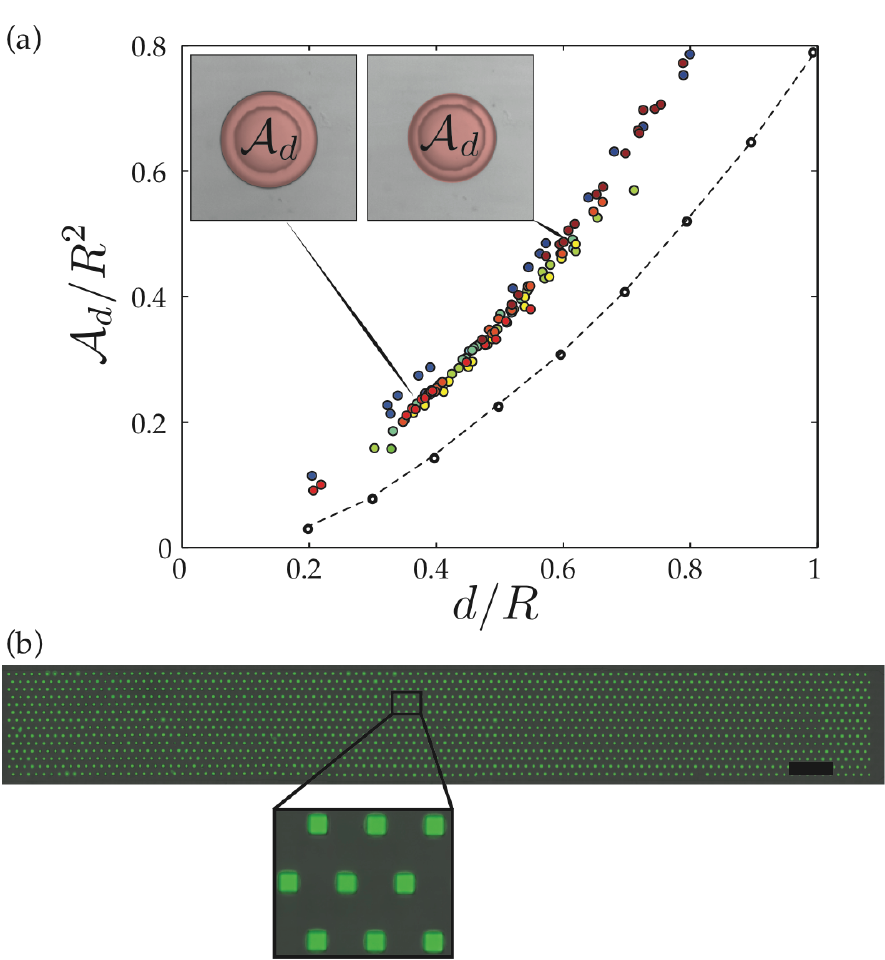}
\caption{(a) Remaining area (dimensionless) as a function of $\overline{d}=d/R$. Dots: experiments, line: theoretical
  prediction. Insets: two droplets of different sizes left on the same anchor ($d=150 \rm{\; \mu m}$) by two different
  drops. (b) An array made up of 1568 individual drops produced through self-digitization. The aqueous
  solution contains fluorescein to aid in the visualisation. We have used square anchors with side $d=130\;\mu\rm m$ in a channel of
  height $h=35\;\mu \rm m$. Scale bar: 2 mm.}
\label{fig:fig4}
\end{center}
\end{figure}

%What is the area left on the anchor?
In many situations it is important to determine the volume of the trapped fluid in the anchor, for example if this is
used to observe biological samples. This is equivalent to predicting, in our 2D view, the projected area $\mathcal
A_{d}$ of the droplet left on the anchor after breakup. We expect the area to depend both on the diameter of the trap $d$ and on the droplet radius $R$. The dependence of $\mathcal A_{d}$ on $d$ is dictated by the the boundary condition that the trap imposes on the droplet shape. The dependence on $R$ comes from the
geometry of the break-up: we experimentally and numerically observe that, for a given $d$, the pinch-off location increases with $R$,
occuring further downstream of the trap for larger drops. Rescaling all lengths by $R$, we show the evolution of the
dimensionless area $\mathcal A_{d}/R^2$ as a function of $\overline{d}=d/R$ for both experimental results and
theoretical simulations on Fig.~\ref{fig:fig4}a. Note that a perfect agreement cannot be expected between the
theoretical and experimental results since the model is 2D, whereas 3D effects are present close to the anchor. Still,
the trends of the two curves are identical.

%why the method is nice
We now examine the practical implications of the demonstrated method.  A droplet breaks on an anchor as long as the
outer flow velocity satisfies $U>U^\star$. In this sense, one does not need a precise flow control to produce a droplet:
pushing the outer flow using a hand-held syringe is sufficient to break droplets. Also, the only relevant viscosity
coming into play is the viscosity $\mu$ of the outer fluid: since $U^\star$ does not depend on the inner fluid
viscosity, the singularity occurs at the same time for droplets of a given volume, so that there is no difference
between breaking droplets of a viscous fluid such as FC-70 ($\mu=24\;\rm cP$) or breaking droplets of HFE-7500 ($\mu =
0.77 \rm\; cP$). The volume  left on the trap is dependent on the geometries of the trap and that of the drop: the sole
important dimensionless parameter is $\overline d$. Therefore, whatever the fluids used, their surface tensions,
viscosities, two drops of the same volume will break on an anchor of diameter $d$ into droplets of the same size,
$\mathcal{A}_{d}$, that depends solely on $d$ and $R$.

%%%Figure 4

%We can do better than 1 anchor: extend the method to 2D arrays of traps
The breakup of a single droplet can be generalized to an array of traps of any dimension, as shown in
Fig.~\ref{fig:fig4}b that shows an array of $112\times14$ drops. In this experiment an aqueous ``puddle'' initially
fills the chamber and is then pushed by a flow of a wetting oil. The drops, which are produced at 10 Hz, detach as the
front passes the anchors. Again, the interface connecting two anchors is decribed by an elastica equation, albeit with
different boundary conditions than the single anchor, and breaks when the enclosed volume is reduced below a critical
value. 

This protocol can directly lead to applications to bio-assays, for instance by working with a suspension of cells in the
aqueous phase. The droplet array recalls the multiwell plate format, where each drop is analogous to a well
encapsulating a cell population that is independent from its neighbors and can be continuously monitored. The oil phase
can then serve to control the gas exchange with the drops, for example to oxygenate or de-oxygenate red blood cells, as
shown previously~\cite{abbyad-dangla}.

The authors thank Caroline Frot for help with the micro-fabrication and David Gonzalez-Rodriguez for helpful
discussions. The research leading to these results has received funding from the European Research Council under the
European Union’s Seventh Framework Programme (FP7/2007- 2013)/ERC Grant agreement no. 278248 ‘MULTICELL’ for Ecole
Polytechnique and ERC agreement no. 280117 `SIMCOMICS' for EPFL.

\bibliography{droplets.bib}

\end{document}